\UseRawInputEncoding
\documentclass[sigconf]{acmart}

\usepackage{amsmath,amsfonts,bm}
\usepackage{microtype}
\usepackage{graphicx}
\usepackage{footmisc}
\usepackage{diagbox}
\usepackage{enumitem}
\usepackage{subfigure}
\usepackage{multirow}

\usepackage{algorithm}% http://ctan.org/pkg/algorithms
\usepackage[noend]{algpseudocode}% http://ctan.org/pkg/algorithmicx

\usepackage{booktabs} % for professional tables

\usepackage{hyperref}

\usepackage{lipsum}

\AtBeginDocument{%
  \providecommand\BibTeX{{%
    \normalfont B\kern-0.5em{\scshape i\kern-0.25em b}\kern-0.8em\TeX}}}

\setcopyright{acmcopyright}
\copyrightyear{2021}
\acmYear{2021}
\setcopyright{acmcopyright}\acmConference[SIGIR '21]{Proceedings of the 44th International ACM SIGIR Conference on Research and Development in Information Retrieval}{July 11--15, 2021}{Virtual Event, Canada}
\acmBooktitle{Proceedings of the 44th International ACM SIGIR Conference on Research and Development in Information Retrieval (SIGIR '21), July 11--15, 2021, Virtual Event, Canada}
\acmPrice{15.00}
\acmDOI{10.1145/3404835.3462976}
\acmISBN{978-1-4503-8037-9/21/07}

\begin{document}

\title{ScaleFreeCTR: MixCache-based Distributed Training System for CTR Models with Huge Embedding Table}
\author{Huifeng Guo{$^\dagger$$^*$}, Wei Guo{$^\dagger$}, Yong Gao, Ruiming Tang$^*$, Xiuqiang He, Wenzhi Liu}

\email{{huifeng.guo,guowei76,neeson.gaoyong,tangruiming,hexiuqiang1,rogy.liu}@huawei.com}
\affiliation{%
  \institution{
  Huawei Noah's Ark Lab
  }
  \country{China}
}
% \authornote[1]

% \affiliation{%
%   \institution{Institute of Anonymous}
%   \streetaddress{X}
%   \city{X}
%   \state{X-State}
%   \country{X-Country}
%   \postcode{0000}
\renewcommand{\shortauthors}{Huifeng Guo and Wei Guo, et al.}

\begin{abstract}
Because of the superior feature representation ability of deep learning, various deep Click-Through Rate (CTR) models are deployed in the commercial systems by industrial companies. To achieve better performance, it is necessary to train the deep CTR models on huge volume of training data efficiently, which makes speeding up the training process an essential problem.
Different from the models with dense training data, the training data for CTR models is usually high-dimensional and sparse.
To transform the high-dimensional sparse input into low-dimensional dense real-value vectors, almost all deep CTR models adopt the embedding layer, which easily reaches hundreds of GB or even TB.
Since a single GPU cannot afford to accommodate all the embedding parameters, when performing distributed training, it is not reasonable to conduct the data-parallelism only.
Therefore, existing distributed training platforms for recommendation adopt model-parallelism. Specifically,  they use CPU (Host) memory of servers to maintain and update the embedding parameters and utilize GPU worker to conduct forward and backward computations.
Unfortunately, these platforms suffer from two bottlenecks: (1) the latency of pull \& push operations between Host and GPU; (2) parameters update and synchronization in the CPU servers. To address such bottlenecks, in this paper, we propose the ScaleFreeCTR: a MixCache-based distributed training system for CTR models. Specifically, in SFCTR, we also store huge embedding table in CPU but utilize GPU instead of CPU to conduct embedding synchronization efficiently. To reduce the latency of data transfer between both GPU-Host and GPU-GPU, the MixCache mechanism and Virtual Sparse Id operation are proposed.
Comprehensive experiments are conducted to demonstrate the effectiveness and efficiency of SFCTR. In addition, our system will be open-source based on MindSpore\footnote{\noindent{MindSpore. https://www.mindspore.cn/, 2020.}} in the near future.

% \blfootnote
\noindent\let\thefootnote\relax\footnotetext{$^\dagger$Contribute to this work equally.}
\footnotetext{$^*$Huifeng Guo and Ruiming Tang are the corresponding authors.}
\end{abstract}

\begin{CCSXML}
<ccs2012>
<concept>
<concept_id>10010147.10010919.10010172</concept_id>
<concept_desc>Computing methodologies~Distributed algorithms</concept_desc>
<concept_significance>300</concept_significance>
</concept>
<concept>
<concept_id>10010147.10010257</concept_id>
<concept_desc>Computing methodologies~Machine learning</concept_desc>
<concept_significance>300</concept_significance>
</concept>
<concept>
<concept_id>10002951.10003317</concept_id>
<concept_desc>Information systems~Information retrieval</concept_desc>
<concept_significance>500</concept_significance>
</concept>
</ccs2012>
\end{CCSXML}

\ccsdesc[500]{Computing methodologies~Distributed algorithms}
\ccsdesc[500]{Computing methodologies~Machine learning}
\ccsdesc[500]{Information systems~Information retrieval}
\keywords{CTR Prediction, Recommendation, Distributed Training System}


\maketitle

\section{Introduction}
To alleviate the problem of information explosion, recommender systems are widely deployed to provide personalized information filtering in online information services, such as web search, news recommendation, and online advertising. In recommender systems, Click-Through Rate (CTR) prediction is a crucial task, which is to estimate the probability that a user will click on a recommended item under a specific context, so that recommendation decisions can be made based on the predicted CTR values~\cite{ftrl,din,wide_deep,deepfm,baidups,gbdt_facebook}.
Due to the superior performance of feature representation in computer vision~\cite{resnet} and natural language processing~\cite{bert}, deep learning techniques attract the attention of recommendation community. Therefore, industrial companies propose various deep CTR models and deploy them in their commercial systems, such as Wide \& Deep~\cite{wide_deep} in Google Play, DeepFM~\cite{deepfm} in Huawei AppGallery and DIN~\cite{din} in Taobao.
\begin{figure}[!t]
    \centering
    \includegraphics[width=0.48\textwidth]{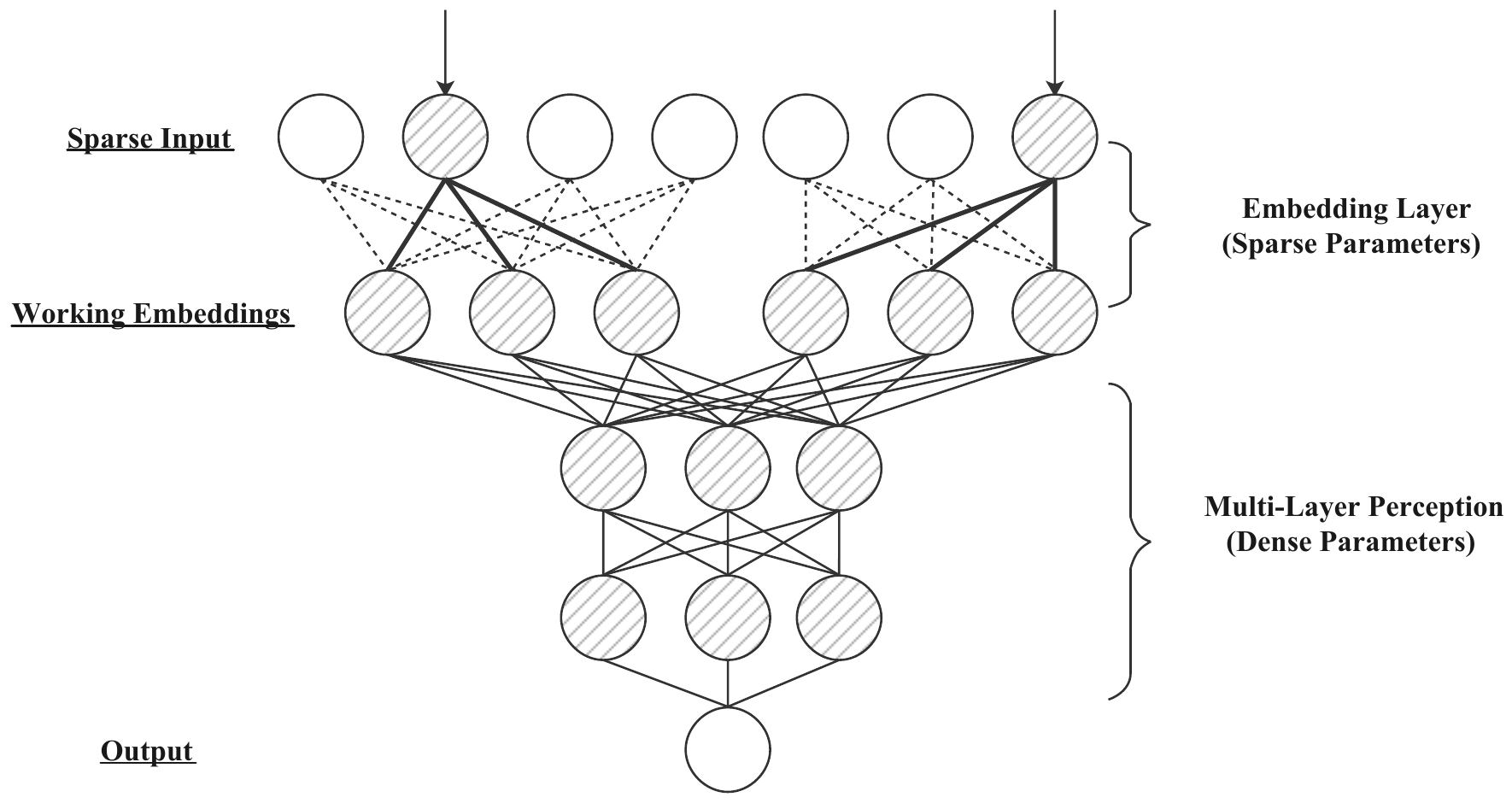}
    \caption{\small{Embedding \& MLP paradigm of CTR models.}}
    % we learn the item embedding $\mathbf{h}_v^k$ with the same mechanism.}
% \vspace{-0.3cm}
    \label{fig:CTR_model}
\end{figure}
To achieve good performance, deep CTR models with complicated network architectures need to be trained on huge volume of training data~\footnote{In industrial scenarios,
tens of billions of training samples are very commonly needed to train a well-performed deep recommendation model~\cite{baidups}.} for several epochs, leading to low training efficiency. Such low training efficiency (namely, long training time) may result in performance degradation when the model is not produced on time and therefore delayed to deploy~\cite{IncreCTR}. Hence, how to improve training efficiency of deep CTR models without hurting model performance is an essential problem in industrial recommender systems. Incremental learning~\cite{IncreCTR,GraphSail,IADM} and distributed training~\cite{xdl,des_tencent,baidups} are two common paradigms to tackle this problem from different perspectives.
Incremental learning is a complement to batch training, which only utilizes the most recent data to update the model. Distributed training utilizes extra computational resources to speed up batch training process.

Many applications, such as computational finance~\cite{dnn_finance}, computer vision~\cite{image_minutes} and natural language processing~\cite{gpt_3}, also need training datasets of billions of instances with TB size.
% In most cases of these applications, the multi-GPU framework is  utilized to accelerate the training process.
To perform distributed training on these applications, data-parallelism, such as the All-Reduce based approach~\cite{hvd,eff_allreduce}, is usually adopted, as the models in such applications are small enough to put in the High Bandwidth Memory (HBM) of a single GPU device (usually no more than 32 GB).
However, different from the above mentioned applications with dense training data, the data of recommendation and advertising is high-dimensional and sparse~\cite{deepfm-end,PinSage,wide_deep}. As shown in Figure~\ref{fig:CTR_model}, existing deep CTR models follow the Embedding \& Multi-Layer Perception (MLP) paradigm. The embedding layer transforms the high-dimensional sparse input into low-dimensional dense real-value vectors~\cite{Fnn}. The sparse features can easily reach a scale of several billions or even trillions~\cite{baidups}, making the parameter size of the embedding layer\footnote{The number of parameters in deep CTR models is heavily concentrated in the embedding layer~\cite{NIS}. We will elaborate the embedding layer in Section~\ref{sec:related-ctr}.} to be hundreds of GB or even TB, which is significantly larger than the HBM of a single GPU device with several orders of magnitude. Therefore, when performing distributed training, it is not a reasonable solution to conduct data-parallelism only, since a single GPU cannot always afford to accommodate all the embedding parameters. Due to this reason, the majority of the existing distributed training frameworks for recommendation consider model-parallelism.
% We analyse the limitations of these existing model-parallelism solutions.

\begin{figure}[!t]
    \centering
    \includegraphics[width=0.48\textwidth]{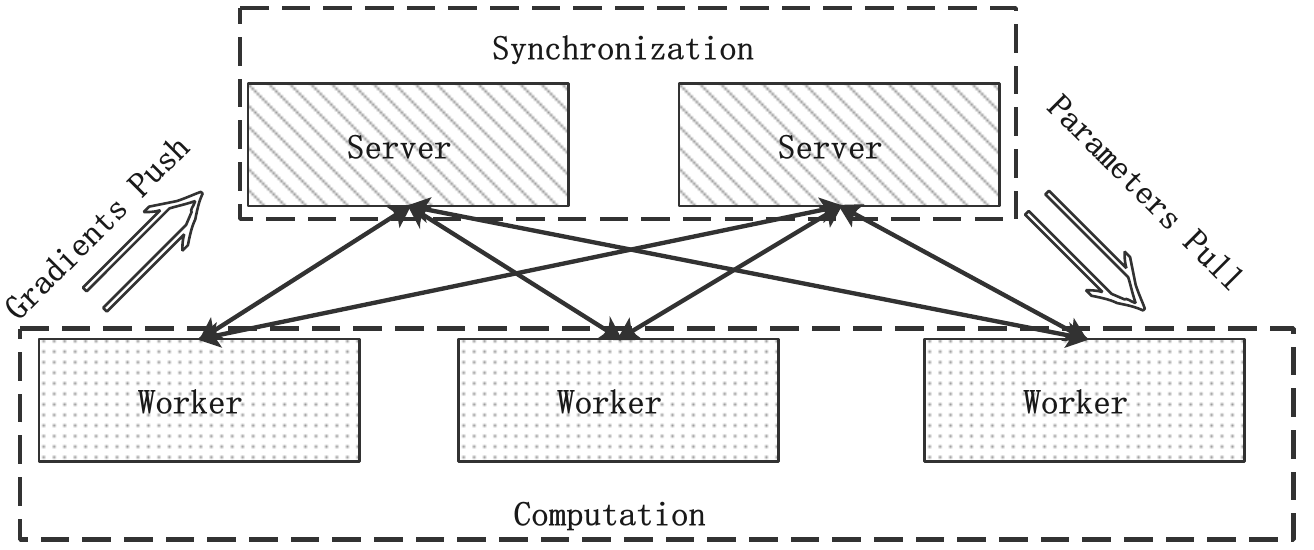}
    \caption{\small{The training process  of parameter server.}}
    % \caption{\small{The training process of parameter server. The upper is the architecture of parameter server and the bottom is an example to illustrate the workflow of parameter server. First, two workers pull the parameters of corresponding features, namely $[1,2]$ and $[2,7]$; second, each worker conducts computations, including forward and backward phases, and obtains corresponding gradients; third, two workers push the gradients to the servers; finally, servers will update the parameters based on the collected gradients.}}

% \vspace{-0.3cm}
    \label{fig:training_process}
\end{figure}

\textbf{Limitations of Existing Model-Parallelism Solutions for Recommendation Models.}
As a classic distributed training framework, Parameter Server (PS)~\cite{ps_muli,xdl,parallax} is widely used to train CTR models with large amount of embedding parameters. Two distinguished roles, namely server and worker, exist in PS mechanism, as shown in Figure~\ref{fig:training_process}. The servers maintain and synchronize model parameters, and the workers perform the forward and backward computations. Specifically, a worker pulls its corresponding parameters from the severs, conducts forward computation to make predictions, computes gradients by back-propagation and finally pushes such gradients to the servers. Deployed in the cluster of high performance devices~\footnote{The high performance device includes Graphical Process Unit (GPU) from NVIDIA, Neural Process Unit (NPU) from Google and Ascend AI Processor from Huawei. Unless otherwise stated, we assume GPUs are equipped when elaborating our work and other works, because GPUs are most widely-used.}, the computation  of recommendation models in workers is very fast and therefore the efficiency of traditional PS usually suffers from these two bottlenecks: (1) pull and push operation between the servers and the workers; (2) parameters synchronization in the servers after receiving gradients from the workers.

To make full use of powerful computation ability and high-speed bandwidth of GPUs, various distributed training systems for deep CTR models are proposed by industrial companies, such as HugeCTR~\footnote{https://github.com/NVIDIA/HugeCTR\label{fnlabel_hugeCTR}} by NVIDIA, DLRM~\cite{DLRM} by Facebook, DES~\cite{des_tencent} by Tencent and HierPS~\cite{baidups} by Baidu. HugeCTR and DLRM devide the parameters of the embedding layer into multiple pieces such that each piece can be stored in the HBM of a single GPU. Due to the limited size of HBM (normally less than 32 GB), we need tens or even hundreds of GPUs to store the embedding parameters, which is expensive. Therefore, the solutions provided by HugeCTR and DLRM are impractical to train deep CTR models with huge embedding table.

To overcome this limitation, DES and HierPS utilize the Host memory to maintain the embedding parameters of large size. More specifically, DES proposes a field-aware partition strategy to reduce communication data among GPUs, while the communication cost between Host-GPUs are not considered to optimize. HierPS introduces a Big-Batch Strategy to cache the working parameters in GPUs to decrease the latency of parameters/gradients transfer between Host-GPUs, but the latency between two such consecutive Big-Batches is not tolerable. Therefore, both of DES and HierPS still suffer from the latency of parameters/gradients transfer between Host-GPUs.

\textbf{Challenges \& contributions.}
From the above observations, we can see that there are two challenges when designing a distributed training framework for deep CTR models. \emph{First}, how to reduce the latency of parameters/gradients transfer between Host-GPUs is a key factor, as a large latency leads to low scalability. \emph{Second}, how to reduce the amount of parameters/gradients transfer between Host-GPUs and GPU-GPU is also important.
%different computation phase, especially the latency of Host-GPU data transfer.
% The longer this latency is, the smaller the scalability is. 2) The Second challenge is to decrease the amount of communication data and enhance the efficiency of communication.

There are two distinguishable characterizations for data in recommendation, which help us to design a distributed training system for deep CTR models and tackle the above mentioned challenges. \emph{First}, working parameters are of small size. Working parameters refer to all the parameters that exist and will be updated in the current batch of training data, including the embedding parameters of all the existing sparse features in the current batch and parameters in MLP. Sparse features existed in the current batch are not many and the parameters in MLP are also limited, which lead to small size of working parameters. \emph{Second}, sparse features follow power-law distribution. That is to say, a very small portion of the sparse features appear in the dataset with high frequencies, which also suggests that there exist many duplicate features in a batch of training data. The first characterization gives the chance to eliminate the latency between Host-GPUs by utilizing the cache mechanism, and the second characterization provides us the opportunity to reduce the amount of parameters/gradients transfer between both Host-GPUs and GPU-GPU by re-organizing the data in the current batch.
% Based on the existing observation~\cite{xdl,baidups} and our analysis, there are two interesting properties in the recommendation data.
% 1) Only a small subset parameters in the embedding table is used and updated for a specific batch of data and the size of dense parameters in the Multi-Layer Perceptron (MLP) of CTR models is small (around a few hundred megabytes). So the working parameters (both the corresponding embedding and MLP parameters in current batch data) can fit in the GPU memory.
% 2) A small set of features are used with higher frequency and some features of batch data are duplicated.
% So it is possible to re-organize the data in current batch, then reduce the amount of communication or transfer of working embedding parameters.

Leveraging the characterizations of small size of working parameters and power-law distribution of features, we propose ScaleFreeCTR (SFCTR for short), to training deep CTR models with huge size of embedding parameters efficiently, in a distributed GPU cluster. SFCTR consists of three modules, i.e., Data-Loader, Host-Manager and GPU-Worker.
These three modules in SFCTR make the following contributions:
\begin{itemize}
    % \item We give a detailed analysis of the existing distributed training platforms for deep CTR models. To address the limitations of existing platforms, we propose SFCTR, consists of Data-Loader, Host-Manager, and GPU-Worker.
    \item SFCTR utilizes Host memory to maintain the huge embedding table and GPU to conduct the synchronization of embedding parameters efficiently\footnote{Embedding table stores the embedding parameters of all the features. We will describe it formally in Section~\ref{sec:related-ctr}.}. To reduce the amount of parameters/gradients transfer between Host-GPUs and GPU-GPU, a Virtual Sparse Id (VSI)  operation is proposed in Data-Loader of SFCTR, removing the duplicate feature embeddings in a batch of data.
    \item To eliminate the latency of parameters/gradients transfer between Host-GPUs,  we propose MixCache, an efficient cache mechanism, in Host-Manager of SFCTR to store the working parameters in GPUs.
    \item A three-stage pipeline, that overlaps Data-Loader, Host-Manager, and the training in GPU-Worker, is proposed to match the inevitable GPU computation.
    \item Comprehensive experiments are conducted on a public benchmark dataset to demonstrate the efficiency of SFCTR, which is 6.9X faster than HugeCTR by NVIDIA and 1.3X faster than the optimized parameter server of MxNet. Ablation studies are also performed to validate the effectiveness and efficiency of VSI operation, MixCache mechanism and three-stage pipeline.
\end{itemize}

 \section{Related Work and Background}

In this section, we discuss related literature of deep CTR models and distributed training platforms for deep CTR models.

\subsection{Deep CTR Models}\label{sec:related-ctr}

CTR prediction models evolves from logistic regression~\cite{ftrl}, factorization machines and its variants~\cite{fm,ffm}, to deep learning models~\cite{deepfm,wide_deep,fgcnn,FiBiNet,deep_cross,autogroup,PAL}.
% According to whether to use embedding, CTR methods can be summarized as two categories: traditional approach and embedding-based approach.
Generalized Logistic Regression (LR) models, such as FTRL~\cite{ftrl}, are widely used in recommender systems, because of their robustness and efficiency.
% However, LR cannot learn and capture feature interactions or nonlinear feature transformation automatically. Therefore, tree-based models, such as GBDT~\cite{gbdt_facebook}, are proposed for CTR prediction.
Recently, due to superior performance of feature representation in computer vision and natural language processing, deep learning techniques attracts more attention in recommendation community.
Most deep CTR models follow the Embedding \& MLP paradigm, as presented in Figure~\ref{fig:CTR_model}. The embedding layer transforms the high-dimensional sparse input (which are usually categorical features\footnote{Features in numerical form are normally transformed into categorical form by bucketing~\cite{pin,fgcnn}.}, such as \texttt{city}, \texttt{gender}, \texttt{user id}) into low-dimensional dense real-value vectors. MLP layers aim to learn the non-linear interactions among features, such as DeepFM~\cite{deepfm}, xDeepFM~\cite{xdeepfm}, DIN~\cite{din}.

% Plenty of deep CTR models are proposed and most of them are embedding-based.
% In CTR prediction, learning sophisticated feature interactions is critical. So different network architectures are proposed to capture the feature interactions implicitly and explicitly, such as FM component in DeepFM~\cite{deepfm}, CIN in xDeepFM~\cite{xdeepfm}, attention mechanism in DIN~\cite{din}.

% According to the paradigm of how to model feature interactions, we can summarize the existing models into three
% categories:
% Only-Implicit: Only modeling implicit feature interactions directly based on the learnable embeddings, such as FNN~\cite{Fnn}.
% Explicit-Direct: Modeling both explicit and implicit feature interactions. Explicit and implicit interactions are modeled in parallel through product-based component and MLP respectively. Noted that the explicit interactions are fed into the final prediction without going through MLP. Wide \& Deep~\cite{wide_deep}, DeepFM~\cite{deepfm}, AFM~\cite{afm}, DCN~\cite{deep_cross} and AutoFIS~\cite{autofis} are such instances.
% Explicit-MLP: Modeling both explicit and implicit feature interactions. Explicit interactions are modeled via a special network architecture and then connected to MLP. Finally the output of MLP is served as the final prediction. DIN~\cite{din}, xDeepFM~\cite{xdeepfm}, PIN~\cite{pin}, FGCNN~\cite{fgcnn}, FiBiNET~\cite{FiBiNet} and AutoGroup~\cite{autogroup} fall into this category.

\textbf{Why deep CTR models are of large size.}
As stated earlier, all the features that serve as input for deep CTR models are assumed to be in  categorical form, therefore they have to be encoded into real-value to apply gradient-based optimization methods. One-hot encoding is an approach to encode a categorical feature into a sparse binary vector, which all bits are 0's except for just one bit being 1. For example, the one-hot encoding of a categorical feature with index $i$ is $s_{i}=(0, 0, ..., 1, ..., 0)^T \in \mathbb{R}^{n}$, where $n$ is the number of features in all. The $i$-th of $s_i$ is 1 while other bits are 0's. If followed up a matrix multiplication, we have a low-dimensional representation of the categorical features. That is to say, given a dense matrix $E \in \mathbb{R}^{n\times m}$, $s_{i}^{T}\times E$ is the $m$-dimensional dense real-value vector for feature with index $i$. In this way, a categorical feature is transformed into a low-dimensional dense vector.

However, due to the fact that there are often billions or even trillions of categorical features (i.e., $n$ is very large), the one-hot vector can be extremely high-dimensional. Therefore nowadays it is more usual to use the look-up embedding as a mathematically equivalent alternative. For instance, for the feature with index $i$, it directly looks up the $i$-th row from the dense matrix $E$. The matrix $E$ is often referred to as \emph{embedding table}. As $n$ is usually very huge (in billion- or trillion-scale), the size of embedding table may take hundreds of GB or even TB to hold. Whereas, the parameter size in MLP is relatively much smaller. Therefore, deep CTR models with large embedding table are of large size, and the parameters in the embedding table take the majority of the model size.

\subsection{Distributed Training Platforms for Deep CTR Models}
Investigating advance parallel mechanism among the workers and servers in distributed training systems is important~\cite{ps_muli}.
Distributed training process includes two main phases, namely, computation and parameter synchronization.
% Based on the cluster including workers and servers, the training process of deep CTR models has two main phases: computation and parameter synchronization.
In computation phase, the loss and gradients of parameters are computed according the input data (including both features and labels) in workers. In the synchronization phase, the gradients are collected and adopted to update the maintained parameters by servers. As summarized in ~\cite{baidups}, there are three typical synchronization patterns: (1) Bulk Synchronous Parallel (BSP)~\cite{bsp}, which strictly synchronizes updates from all the workers; (2) Stale Synchronous Parallel (SSP)~\cite{SSP}, which synchronizes results from faster workers; (3) Asynchronous Parallel (ASP), which does not require any synchronization on the gradients from different workers.
Although ASP and SSP improves training efficiency by reducing synchronization frequency, they suffer from degrading model performance (such as accuracy, convergence rate)~\cite{Tradoff_acc_time}. Aiming to not sacrificing model performance, we focus on comparing the existing distributed training systems with BSP. Our proposed SFCTR is also working with BSP.

We compare the existing distributed training systems for deep CTR models with BSP. To improve the inferior performance of distributed training in Tensorflow~\cite{tensorflow}, MxNet~\cite{mxnet} and Pytorch~\cite{pytorch}, both Horovod~\cite{hvd} and BytePS~\cite{BytePS} are proposed to support different platforms. Horovod speeds up the training of dense models based on its efficient and elegant implementation of Ring-AllReduce mechanism. BytePS further improves synchronizing parameters in different layers by prioritized scheduling, i.e., optimizing the order of different layers to synchronize parameters during the backward-propagation and the forward-computation phase.
Recently, to training deep CTR models with a large embedding table, various systems are proposed, such as HugeCTR~\textsuperscript{\ref{fnlabel_hugeCTR}}, DLRM~\cite{DLRM}, DES~\cite{des_tencent} and HierPS~\cite{baidups}. The comparison of SFCTR with the most relevant platforms are summarized in Table~\ref{tab:related}.

\textbf{Relationship with related work:}
% The difference is mainly in two aspects: 1) the mechanism of maintaining host embedding and the working embedding; 2) the communication method between GPUs when synchronize the parameters and gradients of working embedding parameters.
\emph{First}, HugeCTR and DLRM do not design to hold embedding table in the Host memory, therefore models with very large embedding tables (hundreds of GB or TB) are not supported.
\emph{Second}, as DES, HierPS and SFCTR support storing embedding in Host, the latency of parameters/gradients transfer between Host-GPUs raises an issue. DES does not specifically optimize the latency between Host-GPUs. HierPS adopts the Big-Batch strategy to reduce such latency, but there still exists intolerable latency between such two consecutive Big-Batches. To eliminate this latency, SFCTR is equipped with the MixCache, an efficient cache mechanism, to store the working parameters in GPUs, such that much parameters/gradients transfer between Host-GPUs can be avoided.
\begin{table}[!t]
\caption{\small{The comparison between different approaches.}}
\centering
\resizebox{0.5\columnwidth}{!}{%
\begin{tabular}{l|ccc}
\hline \hline
\multirow{2}{*}{} & Embedding &Host-GPUs  \\
& in Host &Latency \\ \hline
HugeCTR &$\times$ & Not Involve \\
DLRM & $\times$ &Not Involve\\
DES & $\surd$ & Large  \\
HierPS  &$\surd$ & Small\\
SFCTR  &$\surd$ & Hiding \\ \hline
\hline
\end{tabular}
}

\label{tab:related}
\end{table}

 \section{Overview of ScaleFreeCTR}\label{sec:overall}

\begin{algorithm}
  \caption{The workflow of ScaleFreeCTR.}
  \label{alg:ScaleFreeCTR}
  \begin{algorithmic}[1]
   \While{not converge}
   \State  batch $\gets$ get\_batch();
    \State global\_id, virtual\_id $\gets$  virtual\_sparse\_id(batch);
    \State working, noworking $\gets$ manager\_get(global\_id);
    \State push\_parameters\_to\_cache(working);
    \State pull\_parameters\_to\_host(noworking);
    \State mini\_batches, local\_id $\gets$ shard(global\_id, virtual\_id);
    \For{ $i \in [1, \#GPU] $}
    \State  common\_embed $\gets$ gather\_cache(local\_id);
      \State common\_embed = AllReduce(common\_embed);
       \State batch\_embed\_i $\gets$ gather(global\_embed, mini\_batch\_i);
       \State grad\_i $\gets$ forward\_backward(loss\_i);
       \State grad\_i $\gets$ grad\_synchronize(grad\_i);
       \State update\_sparse(grad\_i);
    \EndFor
   \EndWhile
  \end{algorithmic}
\end{algorithm}
% \subsection{Workflow}
\begin{figure}[!t]
    \centering
    \setlength{\belowcaptionskip}{-0.5cm}
    \includegraphics[width=0.42\textwidth]{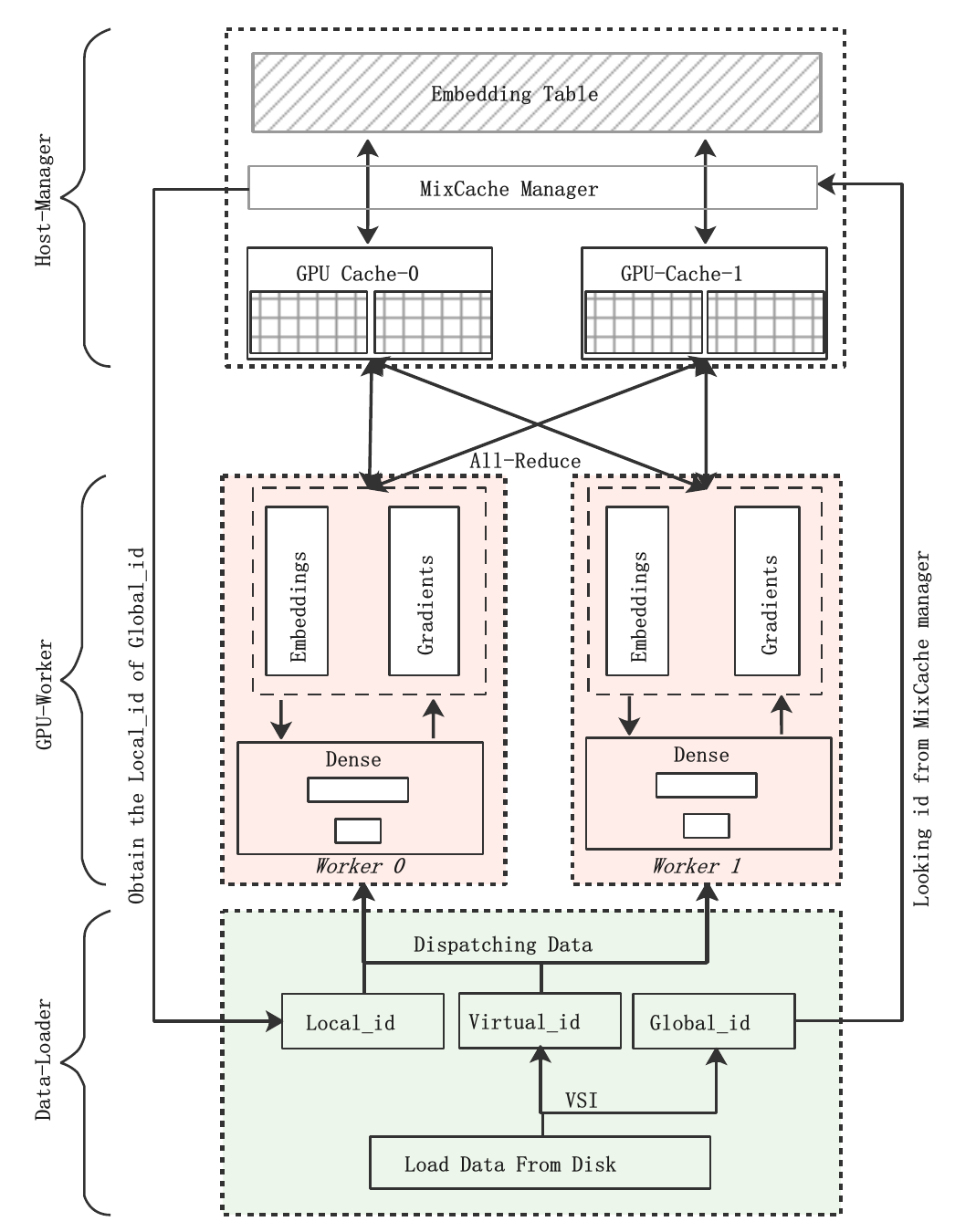}
    \caption{\small{The architecture of ScaleFreeCTR, including Data-Loader  (line 2-3 in Aglorithm~\ref{alg:ScaleFreeCTR}, as presented in Section~\ref{sec:DL}), Host-Manager  (line 4-7 in Aglorithm~\ref{alg:ScaleFreeCTR}, as elaborated in Section~\ref{sec:HM}) and GPU-Worker (line 9-14 in Aglorithm~\ref{alg:ScaleFreeCTR}, as described in Section~\ref{sec:GPU-train}). }}
    % we learn the item embedding $\mathbf{h}_v^k$ with the same mechanism.}
% \vspace{-0.3cm}
    \label{fig:workflow}
\end{figure}

In this section, we overview the architecture and workflow of SFCTR in high-level. As presented in Figure~\ref{fig:workflow}, SFCTR consists of three modules, i.e., Data-Loader, Host-Manager and GPU-Worker (the details are presented in Section~\ref{sec:DL}, Section~\ref{sec:HM} and Section~\ref{sec:GPU-train}, respectively).
At first, we introduce the workflow of SFCTR according to Aglorithm~\ref{alg:ScaleFreeCTR}.
\begin{itemize}
    \item The line 2-3 in Aglorithm~\ref{alg:ScaleFreeCTR} is the \textbf{Data-Loader}, which is responsible for loading data from disk to main memory. To reduce the amount of parameters/gradients transfer between Host-GPUs and GPU-GPU, a Virtual Sparse Id (VSI) operation is proposed. To construct a uniform representation of the working embedding parameters in the current batch, VSI operation removes duplicated feature ids in a batch of data and keeps only one copy of feature ids which is indexed by \texttt{global\_id}. For each instance in the current batch, VSI represents a feature by \texttt{virtual\_id}, which serves as a pointer to find the corresponding \texttt{global\_id} (line 3).
    \item The \textbf{Host-Manager} (line 4-7 in Aglorithm~\ref{alg:ScaleFreeCTR}) maintains embedding parameters in Host memory and transfers the working embedding parameters in a batch of data from Host to the corresponding GPUs. To eliminate such latency of parameters/gradients transfer between Host-GPUs, we propose MixCache\footnote{The ``Mix'' means MixCache is a mixing cache consists of GPU and CPU. Specifically, the manager part of MixCache is in the memory of CPU and the buffers of MixCache are in the HBM of GPUs.}, an efficient cache mechanism in the Host-Manager to store the working parameters in GPUs. MixCache allocates a cache buffer for each GPU, and stores working embedding parameters in each of such cache buffers in a non-overlap manner. MixCache also functions to update cache buffers. More specifically, MixCache checks which embedding parameters are needed for the next batch (referred as \texttt{working} in line 4), and predicts which embedding parameters are not needed for a short future (referred as \texttt{noworking} in line 4). MixCache pushes the \texttt{working} from Host to cache buffers (line 5) and pulls \texttt{noworking} from cache buffers to Host if the cache buffers are full (line 6). Next, Host-Manager dispatches training data to individual GPUs, and identifies \texttt{local\_id} of each feature in the local GPU (line 7).
    \item The \textbf{GPU-Worker} completes the training process in GPUs, including embedding table lookup (line 9-11), forward prediction and backward gradients computation (line 12), MLP parameters and working embedding parameters updating (line 13-14).  Host-Manager and GPU-Worker act in producer-consumer fashion, namely, Host Manager prepares embedding parameters of a batch of data while GPU-Worker consumes the batch of data and updates parameters.
\end{itemize}

\textbf{Three-stage pipeline}:
\begin{figure}[!t]
    \centering
    \includegraphics[width=0.48\textwidth]{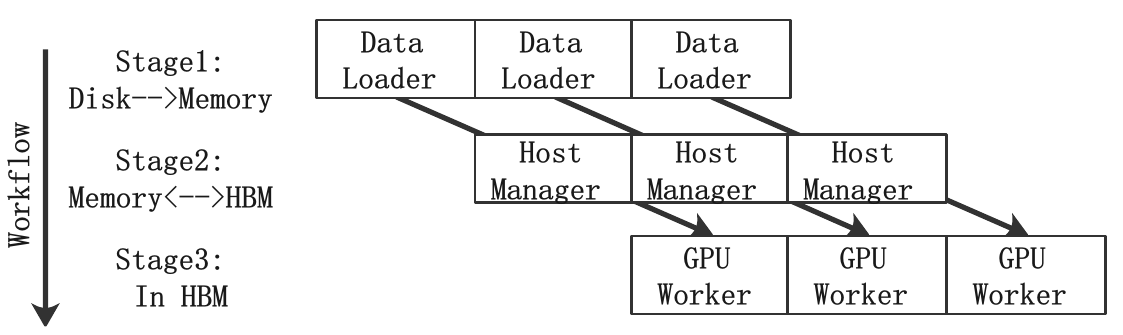}
    \caption{\small{The three-stage pipeline. }}
    % we learn the item embedding $\mathbf{h}_v^k$ with the same mechanism.}
% \vspace{-0.3cm}
    \label{fig:pipeline}
\end{figure}
As presented in Figure~\ref{fig:pipeline}, a three-stage pipeline is applied to overlap timeline of different modules in SFCTR using multi-thread technologies. The training process involves three stages: (1) Data-Loader loads data and removes duplicated feature embeddings by keeping \texttt{global\_id} and \texttt{virtual\_id}; (2) Host-Manager pushes the working embedding parameters to the cache buffers of GPUs and pulls the unnecessary embedding parameters to Host; (3) GPU-Worker performs forward/backward calculation and conducts communication on GPUs. The hardware resources needed in the three stages are different: namely, Disk, CPU and GPU, respectively. Therefore, such different hardware resources can be fully utilized, if jobs of these three stages are scheduled to three different thread pools reasonably. The Host-Manager thread receives global batches of data from the Data-Loader thread and pushes the processed data/parameters to the GPU-Worker thread, as illustrated in Figure~\ref{fig:pipeline}. When the GPU-Workers are busy with training on the current batch of data, the Data-Loader thread is loading data and the Host-Manager thread is preparing parameters in cache buffers from the next batch. Therefore, once the training on the current batch is finished, the required embedding parameters of the next batch are already in HBM of GPUs, so that the training on the next batch can be started immediately.

\section{Data-Loader}\label{sec:DL}

The Data-Loader reads batches of training data from disk to Host memory. In datasets for CTR prediction problem, there exist many repeated features across different data samples in batches. Such redundancy brings unnecessary overhead on parameters/gradients transfer between Host-GPUs in Host-Manager and between GPU-GPU in GPU-Worker. To reduce such redundancy, we propose a Virtual Sparse Id (VSI) operation in Data-Loader. To construct a uniform representation of the working embedding parameters in the current batches, VSI operation removes duplicate feature embeddings in a batch of data and keeps only one copy of feature embeddings which are indexed by \texttt{global\_id}. For each instance in the current batch, VSI represents a feature by \texttt{virtual\_id}, which serves as a pointer to find the corresponding \texttt{global\_id}.

\noindent\textbf{Example for the VSI operation in Data-Loader:} There is a batch with two instances: $\{[1,3,2], [2,3,1]\}$, where each number represents a feature index. After eliminating duplicate features, VSI operation keeps their \texttt{global\_id}, sorted by the order of appearance, as $(1,3,2)$. VSI operation represents the two data instances as $\{[0, 1, 2], [2, 1, 0]\}$, where each number is a \texttt{virtual\_id}. $\texttt{virtual\_id}=0$ retrieves the \texttt{global\_id} with index 0, which is $\texttt{global\_id}=1$. As we can see, utilizing \texttt{global\_id} and \texttt{virtual\_id} is able to recover the data with duplicate feature embeddings.

As we will see later, working embedding parameters are distributed stored in the cache buffers across GPUs, therefore training model on one GPU may need embedding parameters in other GPUs. Loading the batches that are trained cross all the GPUs (referred to as ``global batches'')\footnote{Note that, although Data-Loader of a GPU loads global batches to construct a uniform representation of parameters, this GPU still only trains the model on the batch dispatched to it for training.} to the current GPU is helpful to construct a uniform representation of working embedding parameters. Such a uniform representation introduces a more efficient GPU-GPU communication scheme, which is detailed in Section~\ref{sec:GPU-train}. The size of
such uniform representation for working embedding parameters is much less when we use \texttt{global\_id} instead of original feature index. In the above example, the size of this uniform representation is halved if we use \texttt{global\_id}, because in the original format there are six feature embeddings while three of them are duplicate and therefore eliminated. To summarize, VSI operation helps construct a more space-efficient uniform representation of working embedding parameters, and therefore reduces the amount of parameters/gradients transfer between GPU-GPU. As will be discussed in Section~\ref{sec:HM}, the amount of transfer between Host-GPUs is also optimized by VSI operation.

\section{Host-Manager}\label{sec:HM}

Host-Manager maintains embedding parameters in Host memory and transfers the working embedding parameters in a batch of data from Host to GPUs. To eliminate the latency of parameters/gradients transfer between Host-GPUs, MixCache, which is an efficient cache mechanism in Host-Manager, is proposed. MixCache allocates a cache buffer in HBM of each GPU, which stores the working embedding parameters for the next batch of training data in the corresponding GPU. MixCache utilizes a hash function (such as modulo hashing) to divide the features into disjointed groups such that each cache buffer in a GPU is responsible for one such group.

Each cache buffer of a GPU stores working embedding parameters, of which the corresponding features are in its responsibility. As shown in the three-stage pipeline in Section~\ref{sec:overall}, MixCache identifies the working embedding parameters in the next batch of data and transfers them from Host to the cache buffer if they are not in the cache buffer for now, when the current batch of data is trained in GPU. Therefore, when the training of the current batch is finished, GPUs do not need to wait the working embedding parameters to be transferred and can train the next batch without any latency. As can be seen, embedding parameters of a working feature are transferred to cache buffer at most once with VSI operation performed beforehand; otherwise, embedding parameters of a working feature are repeatedly transferred to cache buffer if this feature is duplicated in the next batch. That is to say, VSI operation reduces the amount of parameters/gradient transfer between Host-GPUs, which has been stated in Section~\ref{sec:DL}.

% Host-Manager is responsible for storing sparse parameters and managing the MixCache. It identifies the required parameters from global\_id and prepares the sparse parameters. Meanwhile, the cache buffer in HBM is also managed by the Manager through putting the sparse parameters in correct position in advance of the training process.

% As mentioned before, the parameters are distributed among multiple nodes and each node also partitions the parameters into multiple hash tables. The number of hash tables equals to the number of GPUs on each node. In order to maximize the throughput, each hash table is maintained by a manager thread in a thread pool. Each manager thread receives a part of sparse unique ids in global training batch according to modulo hashing. For the id not appeared in host, the id are inserted into hash table and the corresponding parameters are initialized. Then the required parameters are pushed to cache according to cache manager.

% To speedup the synchronous training process, the cache buffer in GPU memory is maintained to store the working parameters. The parameters required by the current training batch is put in the cache in advance of the training process and the GPU-workers can fetch the parameters directly. The parameters shared by the previous and current training batch is located in the same position, so the updates from previous training batch can be applied to train the current training batch without parameter staleness.

As the cache buffer is of limited size, embedding parameters have to be transferred back to Host if the cache buffer is full and new working embedding parameters of next batches are needed to transfer into the cache buffer. Embedding parameters of a feature can be transferred back to Host under two conditions that have to be satisfied both: first, the embedding parameters of this feature finish updating; second, this feature does not exist in the next batches that we consider. The first condition guarantees the updating of embedding parameters not interrupted by parameters transfer. The second condition prevents the overhead cost by transferring embedding parameters of a feature to the cache buffer immediately after pulling it out.

\noindent\textbf{Example for the MixCache in Host-Manager:}

\begin{itemize}
    \item \textbf{Current features maintained in cache:} Assume the buffer size of GPU cache is $12$ and the following features' embedding parameters are maintained in the cache: $[1,3,2,5,4,6,$ $12,13,14,15,9]$, where is only one slot available.
    \item \textbf{The features which will be used in next batch:}  Three features (with $\texttt{global\_id}=6,7,8$) in the next batch need to be transferred in the cache buffer, where exists only one slot.
    \item  \textbf{Removing and transferring:} Assume features with $\texttt{global\_id}=3,2,5,4$ are used in current batch and do not finish updating, therefore they are not candidates for removal.
    Feature with $\texttt{global\_id}=6$ is not reasonable to be removed from the cache buffer, as it is needed in the next batch. As a result, features with $\texttt{global\_id}=1$ is selected to be replaced by features with $\texttt{global\_id}=7$, feature with $\texttt{global\_id}=8$ is positioned in the available slot, and feature with $\texttt{global\_id}=6$ is untouched. Now, the following features' embedding parameters are maintained in the cache: $[7,3,2,5,4,6,12,13,14,15,9,8]$
\end{itemize}

\section{GPU-Worker Training}\label{sec:GPU-train}

% The distributed training process falls in data-parallel paradigm. Each GPU worker processes its own training batch and the gradient synchronization is performed after backward propagation.
As the embedding parameters in the current training batch may be located in different GPUs, both the embedding parameters in the forward phase and the gradients in the backward phase need to be synchronized across different GPUs.
\subsection{Forward}
\begin{figure}[!t]
    \centering
    \setlength{\belowcaptionskip}{-0.5cm}
    \includegraphics[width=0.48\textwidth]{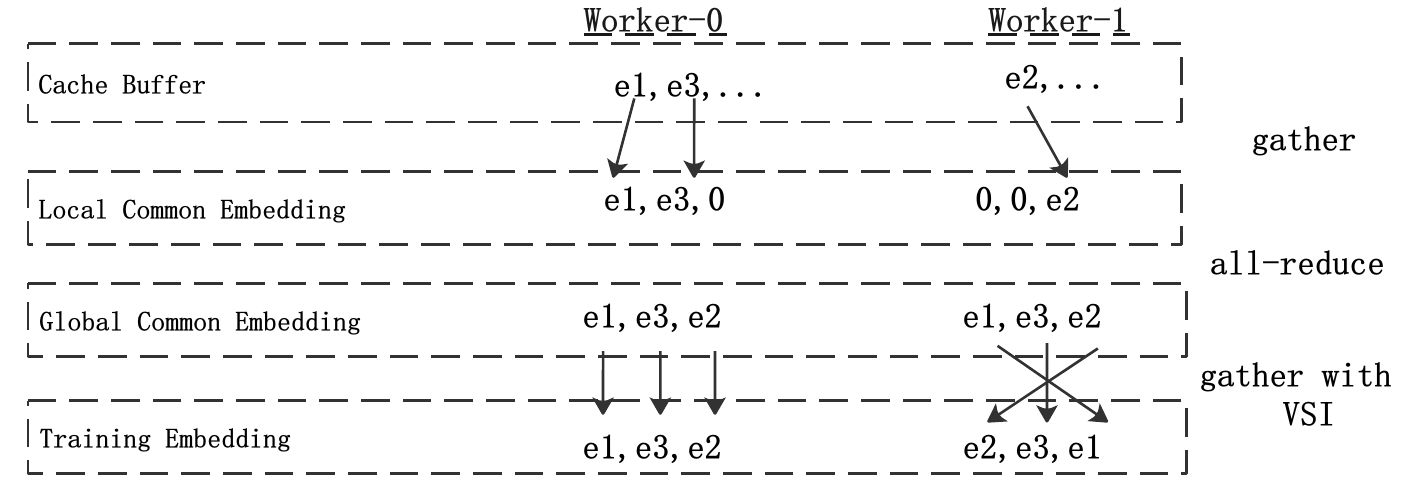}
    \caption{\small{Synchronization of embedding in forward phase.}}
    % we learn the item embedding $\mathbf{h}_v^k$ with the same mechanism.}
% \vspace{-0.3cm}
    \label{fig:forward}
\end{figure}

When synchronizing embedding parameters in the forward phase, every GPU collects the embedding parameters in all GPUs for the current batches of training data.
% All-to-All communication scheme is very common with heavy communication cost, where every GPU needs to communicate with all other ones. To avoid such heavy communication cost,
We choose to adopt all-reduce communication scheme, where every GPU only needs to communicate twice with another two GPUs. To realize the all-reduce communication scheme, a uniform representation of embedding parameters indexed by \texttt{global\_id} is needed in every GPU, which is referred to as Local Common Embedding in Figure~\ref{fig:forward}.

\noindent\textbf{Example of the forward phase in GPU-Worker:} Consistent with the example in Section~\ref{sec:DL}, a global batch of two data instances $\{[1,3,2], [2,3,1]\}$ load for training, where the first instance feeds in Worker-0 and the second one feeds in Worker-1. These two workers know all the \texttt{global\_id} ($(1, 3, 2)$ in this example) in the current global batch of data, realized by Data-Loader, as shown in Figure~\ref{fig:forward}:
\begin{itemize}
    \item \textbf{Gather in Worker-0:} its cache buffer stores $e_1$, $e_3$ and other embeddings, hence, the Local Common Embedding of the current global batch in Worker-0 is $(e_1, e_3, 0)$ ($e_2$ is not in Worker-0 so the position of \texttt{global\_id} 2 is presented with 0 value).
    \item \textbf{Gather in Worker-1:} similarly, the Local Common Embedding of Worker-1 is $(0,0,e_2)$. As the format of the Local Common Embedding is consistent across the workers, all-reduce communication scheme is possible to perform, so that both workers keep the same embedding parameters that are appeared in the current global batch, known as Global Common Embedding in Figure~\ref{fig:forward}.
    \item \textbf{All-reduce:} utilizing \texttt{virtual\_id} of each feature, the original training data with feature embeddings are recovered, as $[e_1, e_3, e_2]$ in Worker-0 and $[e_2, e_3, e_1]$ in Worker-1.
\end{itemize}

% For embedding synchronization in forward phase, instead of conducting all-to-all communication scheme for each worker, all-reduce communication scheme for common embedding matrix is performed to collect all the required embedding for current training batch. As we mentioned before, each worker receives local ids, part of the sparse unique ids in global batch, and the ids in training batch have been transformed to virtual sparse ids. So each worker put the corresponding embedding vectors to the correct position in global embedding matrix, and then all-reduce communication scheme is performed to share the integrated one. Finally, each worker gets embedding matrix for current training batch from common embedding matrix according to virtual sparse ids. Assuming there are two workers and following the example in Figure~\ref{fig:forward}, the parameters of $1,3$ and $2$ belong to worker-1 and worker-0 respectively. Then, the embeddings $[e_1,e_3]$ and $[e_2]$ are gathered from the corresponding cache buffer and the global embedding matrices for worker-1 and worker-2 are generated as $[e1,Null,e3]$ and $[Null,e2,Null]$. After all-reduce communication scheme, each worker deserves the common embedding matrix, namely $[e1,e2,e3]$. Then, each worker can build the embedding matrix for current training batch. Concretely speaking, using  virtual sparse id to gather the embeddings from common embedding matrix.

\subsection{Backward}
\begin{figure}[!t]
    \centering
    \setlength{\belowcaptionskip}{-0.5cm}
    \includegraphics[width=0.48\textwidth]{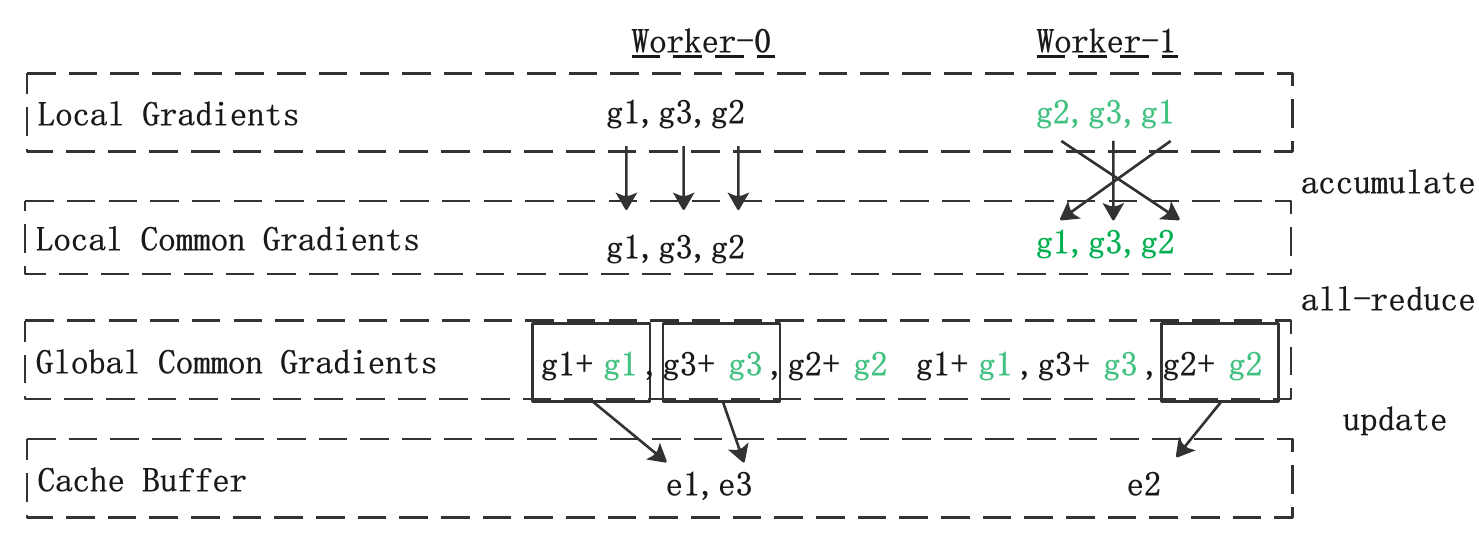}
    \caption{\small{Synchronization and updating of gradients in backward phase.}}
    % we learn the item embedding $\mathbf{h}_v^k$ with the same mechanism.}
% \vspace{-0.3cm}
    \label{fig:backward}
\end{figure}

After backward propagation, the gradients of all the embedding parameters are computed in different workers, known as Local Gradients in Figure~\ref{fig:backward}. Then \texttt{virtual\_id} is utilized to transform Local Gradients to Local Common Gradients to keep a uniform order, i.e, the same as in Local Common Embedding in Figure~\ref{fig:forward}. This transformation is known as unsorted\_segment\_sum in Tensorflow\footnote{https://www.tensorflow.org/api\_docs/python/tf/math/unsorted\_segment\_sum}.
For example in Figure~\ref{fig:backward}, the Local Gradients $(g_1, g_3, g_2)$ and $(g_2, g_3, g_1)$ are accumulated as Local Common Gradients $(g_1,g_3,g_2)$ and $(g_1,g_3,g_2)$ by worker-0 and worker-1 respectively.

After this transformation, the format of the Local Common Gradients is consistent across the workers. So all-reduce communication scheme is possible to perform to collect
the Local Common Gradients of the other GPUs, and the Global Common Gradients are computed by summing such local ones. Finally, each worker uses Global Common Gradients to update the embedding parameters stored in its Cache Buffer. For example, based on the Global Common Gradients, worker-0 updates $e_1,e_3$ and worker-1 updates $e_2$.

% Firstly, the gradients of MLP parameters are synchronized through all-reduce scheme. Then the gradients of sparse parameters are synchronized like embedding synchronization scheme in forward phase. Each worker firstly accumulates the gradients of the sparse parameters (for example in Figure~\ref{fig:backward}) in current training batch to the corresponding position in common embedding matrix, and then perform all-reduce communication. Finally the referenced parameters of global batch in cache is updated
% with the gradients in common embedding matrix. Similarly, we use the same example to illustrate the synchronization of sparse parameters.

 \section{Experiment}
In this section, we conduct extensive experiments to answer the following research questions:
\begin{itemize}
    \item RQ1: How does SFCTR perform compared to other distributed training systems for CTR prediction task?
    \item RQ2: Can VSI operation reduce the amount of communication data and accelerate the training process?
    \item RQ3: Can MixCache mechanism eliminate the latency between Host-GPU without sacrificing the model accuracy?
    \item RQ4: How does the proposed 3-stage pipeline improve the efficiency of SFCTR?
\end{itemize}

\subsection{Setting}

\subsubsection{Environment}
To realize all-reduce communication scheme efficiently, our GPU cluster is connected through InfiniBand which is a high-speed, low latency, low CPU overhead network hardware. Based on InfiniBand network, Remote Direct Memory Access (RDMA) technology is applied to speed-up GPU communication.
Specifically, we evaluate SFCTR on a cluster with four GPU servers connected by a 100Gb RDMA network adaptor. Each server has 2 Intel Xeon Gold-5118 CPUs with 18 cores (36 threads), 8 Tesla V100 GPUs with 32 GB HBM, and approximate 1 TB Host memory. The GPUs in each server are connected with PCIe.

\subsubsection{Dataset}
To make it easy for audiences to reproduce the experiments, we adopt a large-scale real-world public available dataset \emph{Criteo-TB}~\footnote{https://labs.criteo.com/2013/12/download-terabyte-click-logs/} to conduct our experiments and verify the performance of SFCTR.
It consists of 24 days¡¯ consecutive user click logs from Criteo, including 26 categorical features and 13 numerical features. The first column of the dataset is the label indicating whether the ad has been clicked or not.

To investigate the performance of SFCTR to handle different sizes of embedding table, we construct two versions of Criteo-TB, namely 10 GB and 100 GB, through different thresholds (the thresholds of 10 GB and 100 GB are 10 and 1, respectively) to filter low-frequency features (similar to~\cite{pin}). Specifically, the 10 GB (100 GB) contains 33 (330) million features with the size of embedding table $33~(330)~M\times 4~B \times 80~\approx~10~(100)~GB$, where $4~B$ indicates the data type of parameters as float32, $80$ indicates the embedding size. Since we adopt the widely-used Adam as the optimizer, which needs to maintain another two corresponding variables (namely, momentum and velocity), the size of training parameters is three times of the embedding size. Therefore, the parameter size is 300 GB when the embedding table is 100 GB, which is larger than $8 \times 32~GB$ (HBMs of 8 GPUs) and can not be put in GPU memory directly.

\subsubsection{Base Model \& Metric}
We use the widely-used DeepFM~\cite{deepfm} as the base deep CTR model.
In addition, we adopt the training throughput (i.e., number of examples trained per second) and logistic loss (logloss in short) as the evaluation metrics to verify the scalability and accuracy, respectively. Specifically, logloss is defined as:
\begin{equation}
    logloss=\frac{1}{n}\sum_{i=1}^{n}log(1+exp(-y_i\times \hat{y_i}))
\end{equation}
where $n$ is the number of instances, $y_i\in \{0,1\}$ indicates the ground truth of click or not, and $\hat{y_i}\in[0,1]$ is predicted by model.

\subsubsection{Baselines}
We compare the performance of SFCTR with two public available distributed training systems: HugeCTR and Parameter Server (PS in short, based on MxNet~\cite{mxnet}).

HugeCTR is a fully optimized CTR training system that proposed by NVIDIA and achieves better performance than Tensorflow\footnote{\url{https://www.nvidia.cn/content/dam/en-zz/zh_cn/assets/webinars/nov19/HugeCTR_Webinar_1.pdf}}. It is based on the collective communication and only utilizes the HBM of GPUs to store the embedding parameters.

PS utilizes Host memory to store the parameters and transfers the parameters based on the traditional push/pull mechanism.
However, as far as we know, the model training with large embedding in Tensorflow,  MxNet and PyTorch is not well supported, due to the following reasons: (1) there is no officially available PS for PyTorch; (2) the model size in MxNet is limited because of some implementation issues\footnote{\url{https://github.com/apache/incubator-mxnet/issues/17722}}; (3) the throughout of TensorFlow drops sharply when its PS is utilized.
For fair comparison, we use an optimized PS in MxNet with three modifications: (1) we re-implement the parameter server because it does not support the models with huge embedding table; (2) we optimize the training pipeline to overlap the different components of training process; (3) the unique operation is adopted on the batch data of each worker to reduce the communication latency.

In addition, to demonstrate the effectiveness of different components of SFCTR, we also conduct alabtion study for SFCTR.

\subsection{Overall Performance (RQ1)}

 \begin{figure}[!t]
    \centering
    \setlength{\belowcaptionskip}{-0.1cm}
    \includegraphics[width=0.48\textwidth]{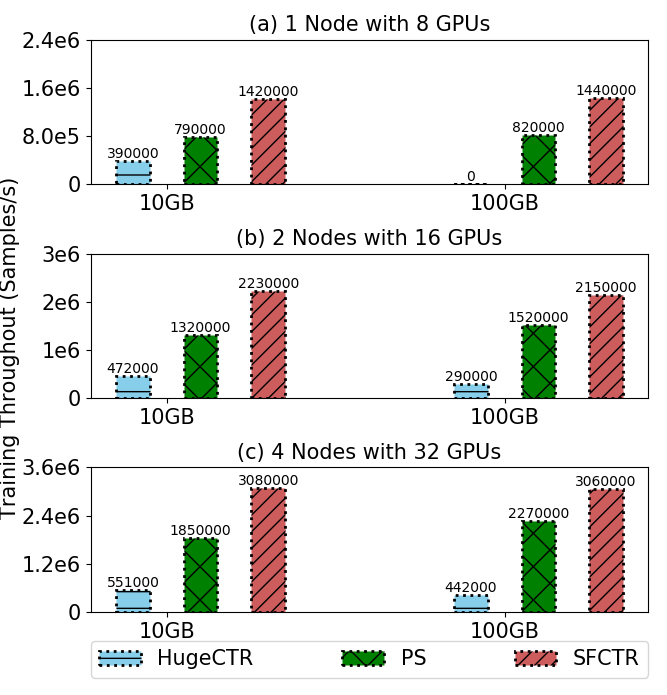}
    \caption{\small{The training throughout comparison of SFCTR, HugeCTR and PS.}}
    % we learn the item embedding $\mathbf{h}_v^k$ with the same mechanism.}
% \vspace{-0.3cm}
    \label{fig:exp:scale}
\end{figure}

To compare the scalability of SFCTR with HugeCTR and PS, we conduct the experiments under three cases with different computational resources: (1) 1 node with 8 GPUs; (2) 2 nodes with 16 GPUs; (3) 4 nodes with 32 GPUs. Figure~\ref{fig:exp:scale} shows the training throughouts of the three compared systems with 10 GB and 100 GB embedding tables (in fact 30 GB and 300 GB parameters, respectively), in the three mentioned cases.
Noted that HugeCTR is not applicable with 300 GB embedding parameters in 8 GPUs because the HBM of GPU(s) cannot afford the corresponding embedding parameters.
We summarize the observations from Figure~\ref{fig:exp:scale} as follows.
\begin{itemize}
    \item Equipped with VSI operation, MixCache mechanism and 3-stage pipeline, SFCTR achieves the highest training throughout under all the settings. The size of embedding parameters does not impact on the thoughout of SFCTR much.
    When 4 nodes with 32 GPUs are available, the acceleration of SFCTR compared to PS (HugeCTR) is 1.6X (5.6X) and 1.3X (6.9X) with 10 GB and 100 GB embedding table, respectively.
    \item The size of a model that can be trained by HugeCTR is limited by the number of cards available. With eight GPUs, HugeCTR cannot even train a model with 100 GB embedding table (300 GB embedding parameters), because HugeCTR only utilizes the HBM of GPUs to store the embeddings.
    \item
    % Both PS and HugeCTR suffer from inferior scalability.
    % Specifically, from single node to 4 nodes, the order of training throughout is PS < HugeCTR < SFCTR.
    As VSI operation reduces the amount of communication data between GPU-GPU, SFCTR is more efficient than HugeCTR. SFCTR is also superior to PS, which thanks to eliminating the latency between Host-GPU by MixCache.
\end{itemize}

%  \begin{figure}[!t]
%     \centering
%     \setlength{\belowcaptionskip}{-0.5cm}
%     \includegraphics[width=0.48\textwidth]{img/scalability-self.png}
%     \caption{\small{Speed up of SFCTR.}}
%     % we learn the item embedding $\mathbf{h}_v^k$ with the same mechanism.}
% % \vspace{-0.3cm}
%     \label{fig:exp:scale}
% \end{figure}

% \begin{table}[!t]
% \caption{\small{The Training Time of Large Model.}}
% \centering
% \resizebox{.9\columnwidth}{!}{%
% \begin{tabular}{l|c|c|c|c}
% \hline \hline
%  & \#GPU=1 &\#GPU=2&\#GPU=4&\#GPU=8 \\ \hline
%  HugeCTR &&&&\\ \hline
% Tensorflow  & & & & \\ \hline
% SFCTR & & & &  \\ \hline
% \hline
% \end{tabular}
% }

% \label{scalability-large}
% \end{table}

% \begin{figure}[!t]
%     \centering
%     \setlength{\belowcaptionskip}{-0.5cm}
%     \includegraphics[width=0.42\textwidth]{Figures/Training Curve of SFCTR.}
%     \caption{\small{Training and test curve of different models on SFCTR, to demonstrate there is no performance loss compared with other platform.}}
%     % we learn the item embedding $\mathbf{h}_v^k$ with the same mechanism.}
% % \vspace{-0.3cm}
%     \label{fig:train-overall}
% \end{figure}

\subsection{The Study of VSI (RQ2)}
\begin{figure}[!t]
    \centering
    \includegraphics[width=0.48\textwidth]{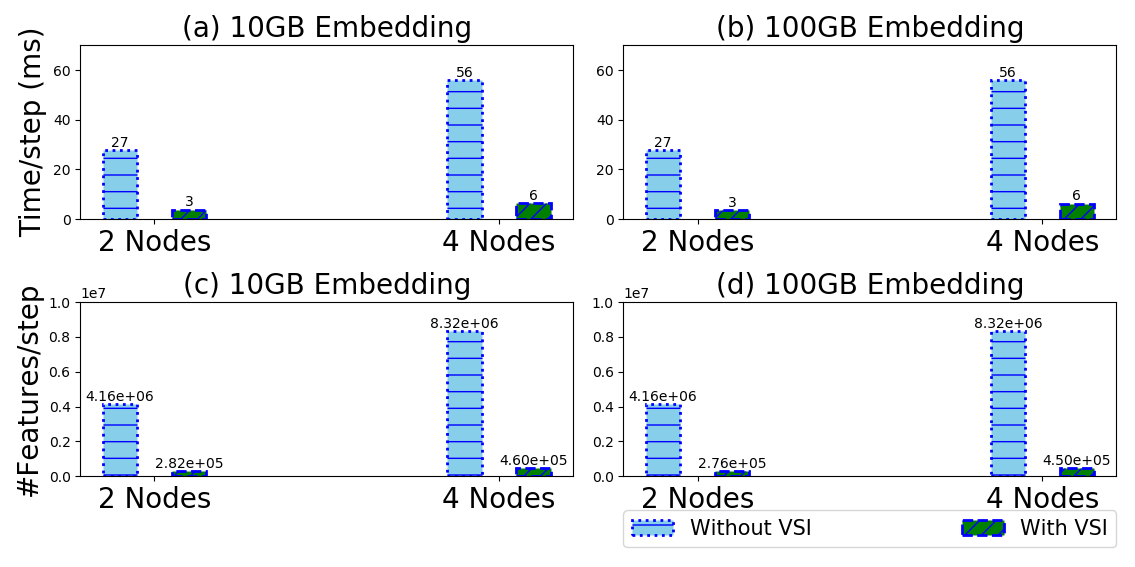}
    \caption{\small{The comparison of data transfer  between using VSI and not using VSI.}}
    % we learn the item embedding $\mathbf{h}_v^k$ with the same mechanism.}
% \vspace{-0.3cm}
    \label{fig:exp:ab_unique}
\end{figure}

Figure~\ref{fig:exp:ab_unique} presents the amount of communication data reduced and the training time saved with the utilization of VSI. Recall that VSI operation sets a global identifier to each feature and eliminates duplicate features in the global batch, which benefits from the following two aspects: (1) it avoids repeatedly transferring duplicate features from Host to cache buffer in GPUs, which reduces the amount of parameters/gradients transfer between Host-GPUs; (2) it helps construct a uniform representation of working embedding parameters which introduces a more efficient GPU-GPU communication scheme. As Figure~\ref{fig:exp:ab_unique} shows, when 4 nodes with 32 GPUs are available, the amount of parameters/gradients transfer between Host-GPU and the communication time between GPU-GPU reduces 94\% and 88\%, respectively, with VSI operation.

% To make better understand of VSI, we give the analysis of communication pattern as follows. Firstly, we define the batch size as B, the number of fields as S, the embedding size as e, the number of GPUs as N, the bandwidth of the allreduce and alltoall algorithm as $B_r$ and $B_a$, and the number of VSI features in a batch as $Num_{VSI}$. For allreduce algorithm, the total communication time is 2 * $Num_{VSI}$ * e * (N - 1)/( N * $B_r$), while for alltoall algothrim, the communication time is S * B * e *  (N - 1)/(N * N * $B_a$). The ratio of communication time between allreduce and alltoall is R=2 * $Num_{VSI}$ * N * $B_a$/($B_r$ *S *B). In the experimental cluster with 2 nodes (N=8 * 2 = 16) connected with InfiniBand connection, $B_r$=10GB/s and $B_a$=2GB/s. For recommendation dataset, e.g. criteo, the repetition rate is more than 90\%, which means $Num_{VSI} < 0.1 * S * B$.
% Therefore, it can be calculated that $R<0.64$, which verifies that the allreduce communication pattern is more efficient than optimized alltoall.
% that utilizing VSI operation can reduce the duplicated features in global batch data.
% Consequently, the average transferred embedding/gradients in each step between GPUs is decreased and the communication time is much smaller than the strategy used in HugeCTR.
% \begin{table}[!t]
% \caption{\small{Effectiveness of Unique-AllReduce Framework.}}
% \centering
% \resizebox{.9\columnwidth}{!}{%
% \begin{tabular}{l|c|c|c|c}
% \hline \hline
%  & \#GPU=1 &\#GPU=2&\#GPU=4&\#GPU=8 \\ \hline
% HugeCTR &&&&\\ \hline
% SFCTR& & & &  \\ \hline
% \hline
% \end{tabular}
% }

% \label{scalability-unique}
% \end{table}
\subsection{The Study of MixCache (RQ3)}
\subsubsection{The cache mechanism} \label{sec:cache-mechanism}
To study the efficiency improvement brought by MixCache, we introduce another two strategies: \textbf{Host} and \textbf{Prefetch}. \textbf{Host} indicates all the embedding parameters are maintained in host memory without pre-fetching parameters in the next batch. \textbf{Prefetch} means prefetch working embedding parameters of next batch before the worker conducts training of current batch, without any guarantee on the consistence of parameters. \textbf{Cache} is the strategy that utilizes the proposed MixCache.

\begin{figure}[!t]
    \centering
    \setlength{\belowcaptionskip}{-0.2cm}
    \includegraphics[width=0.48\textwidth]{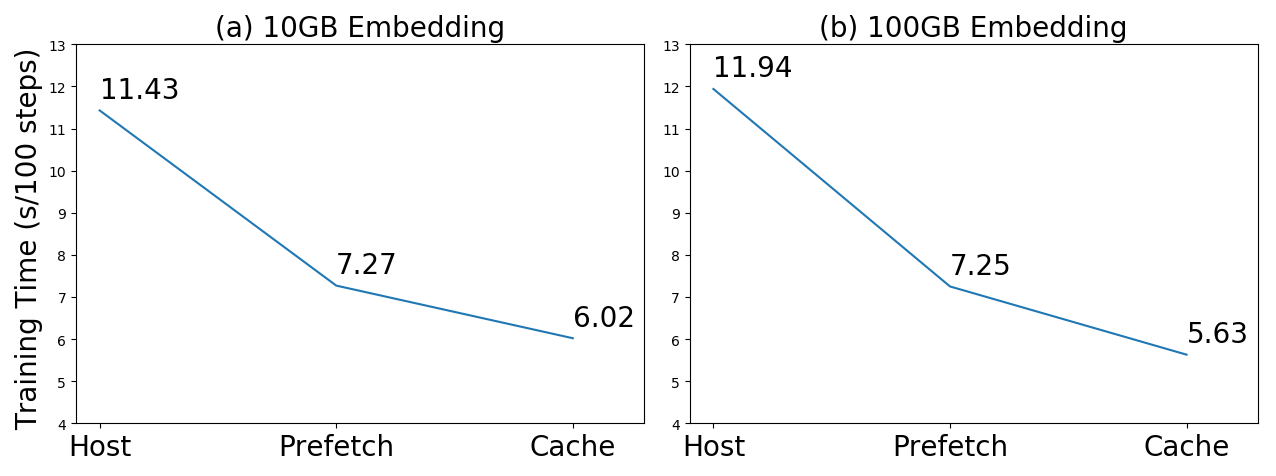}
    \caption{\small{Training time comparison between Host, Prefetch and Cache.}}
    % we learn the item embedding $\mathbf{h}_v^k$ with the same mechanism.}
% \vspace{-0.3cm}
    \label{fig:exp:ab_effci}
\end{figure}

\begin{figure}[!t]
    \centering
    \setlength{\belowcaptionskip}{-0.2cm}
    \includegraphics[width=0.46\textwidth]{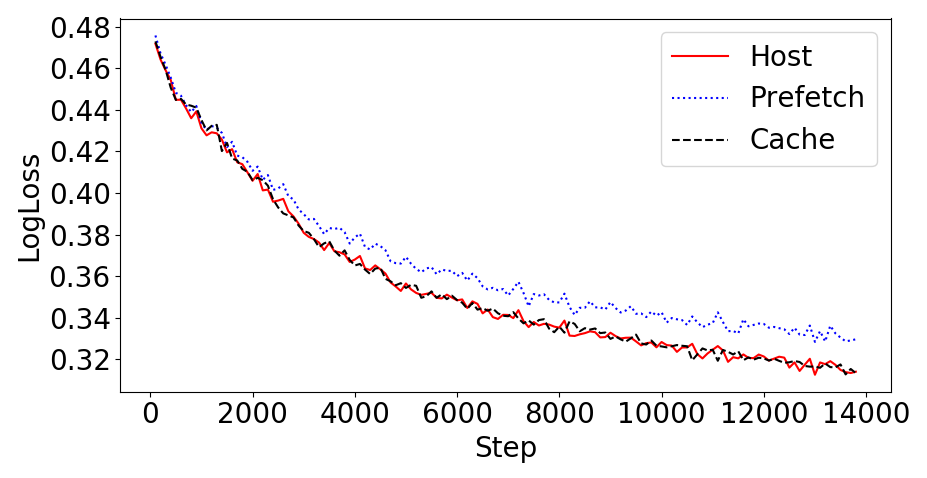}
    \caption{\small{The training curves of Host, Prefetch and Cache strategies.}}
    % we learn the item embedding $\mathbf{h}_v^k$ with the same mechanism.}
% \vspace{-0.3cm}
    \label{fig:exp:ab_curve}
\end{figure}

Figure~\ref{fig:exp:ab_effci} and Figure~\ref{fig:exp:ab_curve} present the training efficiency and training curve of the above mentioned strategies.
% \textbf{Host}, \textbf{Prefetch} and \textbf{Cache}. The Host indicates embedding is maintained in host memory without any pre-fetch strategies, the Prefetch means adding trivial pre-fetch strategy (only pre-fetch working embedding parameters of next batch before worker conducts training of next batch, can not guarantee the consistence of parameters), Cache indicates utilizing the proposed MixCache.
From these figures, we have the following observations:
\begin{itemize}
    \item As presented in Figure~\ref{fig:exp:ab_effci}, to handle both 10 GB and 100 GB embedding tables, Prefetch strategy is more efficient than Host strategy because Prefetch overlaps fetching embedding parameters of the next batch with training the current batch to reduce the overall latency. Moreover, Cache strategy is the most efficient strategy, because the amount of transferring data in Cache strategy is smaller than that in Prefetch, which will be analyzed in~Section~\ref{sec:exp:cache_swap}.
    \item Compared with Host and Cache, the training curve of Prefetch in Figure~\ref{fig:exp:ab_curve} indicates slower convergence, since such simple strategy to prefetch the embedding parameters of the next batch is not able to guarantee the consistency of parameters. Such consistency is promised in both Host and Cache strategies while the Cache strategy implements it in a more efficient way.
    % the training convergence of Cache is faster because the working embedding parameters maintained in cache are up to date. In other word, the MixCache mechanism is able to guarantee the consistence of parameters.
\end{itemize}

% \begin{table}[!t]
% \caption{\small{The efficiency of Pre-fetch strategy.}}
% \centering
% \resizebox{.9\columnwidth}{!}{%
% \begin{tabular}{l|c|c|c|c}
% \hline \hline
%  & \#GPU=1 &\#GPU=2&\#GPU=4&\#GPU=8 \\ \hline
% Vanilla &&&&\\ \hline
% Pre-fetch  & & & & \\ \hline
% \hline
% \end{tabular}
% }

% \label{scalability-cache}
% \end{table}

\subsubsection{The parameter swapping in MixCache}\label{sec:exp:cache_swap}
\begin{figure}[!t]
    \centering
    \includegraphics[width=0.48\textwidth]{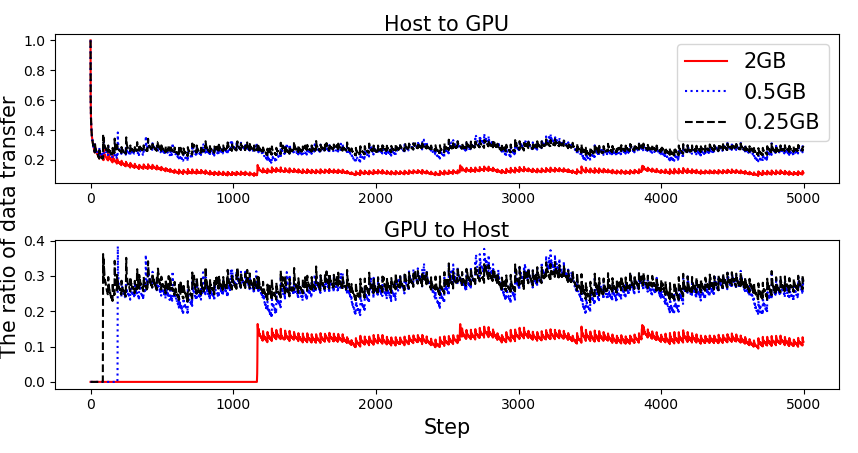}
    \caption{\small{The ratio of data transfer between GPU and Host according to different cache sizes.}}
    % we learn the item embedding $\mathbf{h}_v^k$ with the same mechanism.}
% \vspace{-0.3cm}
    \label{fig:exp:cache_ana}
\end{figure}

To study the impact of the cache size in MixCache, we analyze the ratio of data that need to be transferred between GPU-Host in the current batch, when the cache size is varying in the range of 2 GB, 0.5 GB and 0.25 GB.
% , which is the portion of data need to be transferred in current batch, both from GPU to Host and Host to GPU when cache size is 2GB, 0.5GB and 0.25GB.
The results are presented in Figure~\ref{fig:exp:cache_ana}, where the upper (bottom) half displays the ratio of data transfer from Host to GPU (from GPU to Host) in the different training step. In detail, the red solid, blue dotted, black dashed lines represent results of 2 GB, 0.5 GB and 0.25 GB, respectively. The following observations can be concluded:
\begin{itemize}
    \item A cache with larger cache size makes data transfer from GPU to Host starting later, because a larger cache is able to maintain more embedding parameters. Specifically, a 2 GB-size cache is able to delay the data transfer from GPU to Host after more than 1000 training steps.
    \item The ratio of data transfer in each step is smaller if cache size is larger. This is because the larger cache is able to maintain more embedding parameters of high-frequency features. In particular, the average ratio of data transfer between GPU-Host is around 12\%, 27\% and 29\%, when the cache size is 2 GB, 0.5 GB and 0.25 GB, respectively.
\end{itemize}
Moreover, the analysis of parameter swapping in MixCache also explains why the amount of transferring data in Cache strategy is smaller than that in Prefetch (as in Section~\ref{sec:cache-mechanism}).

\subsection{3-Stage Pipeline (RQ4)}
\begin{figure}[!t]
    \centering
    \setlength{\belowcaptionskip}{-0.5cm}
    \includegraphics[width=0.45\textwidth]{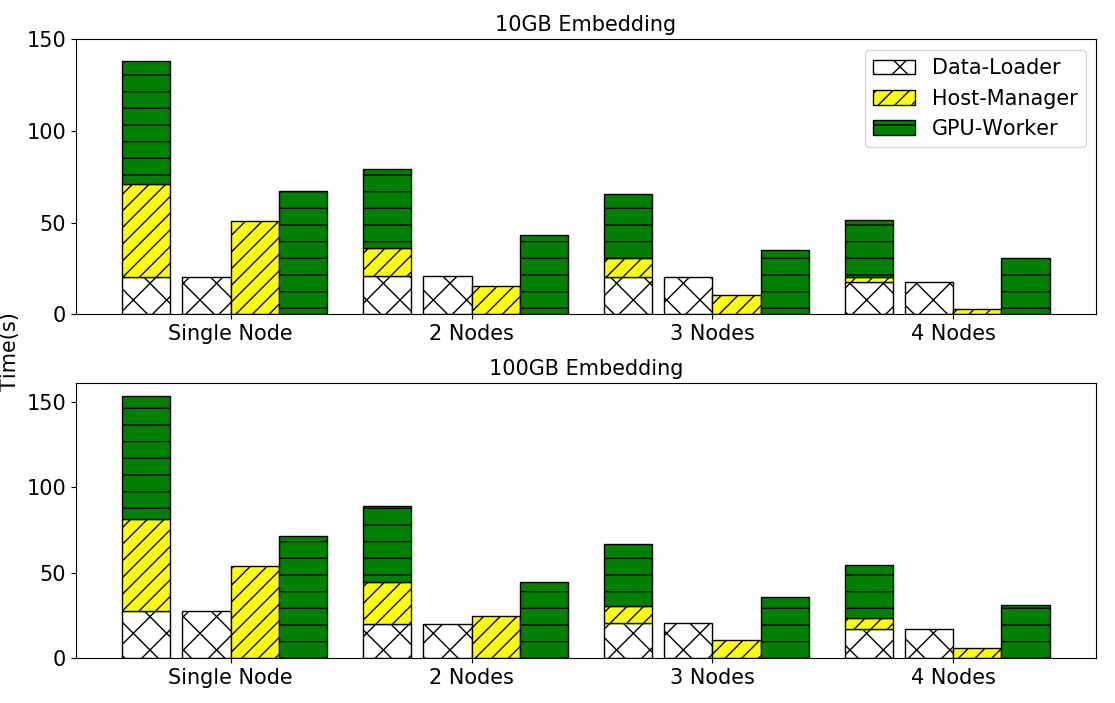}
    \caption{\small{The training time in different stages of SFCTR.}}
    % we learn the item embedding $\mathbf{h}_v^k$ with the same mechanism.}
% \vspace{-0.5cm}
    \label{fig:exp:3-stage}
\end{figure}

In this section, we investigate how effective our proposed 3-stage pipeline is in terms of training time saved. In Figure~\ref{fig:exp:3-stage}, the upper and bottom half present the training time of 100 million samples with 10 GB and 100 GB embedding table, respectively. There are four groups of histograms, namely the training time of single node, 2 nodes, 3 nodes and 4 nodes. Each group contains the training time of different stages.

We can observe that 3-stage pipeline is able to maximize the utilization of GPU worker. As the training time in ``GPU-Worker'' is the largest in all the three stages, such that GPUs can keep computing without interrupting if the other stages can be scheduled reasonably by our pipeline. As a result, our proposed 3-stage pipeline saves the training time significantly.

For example, when we use single node equipped with 3-stage pipeline to train the model with 100 GB embedding, such three stages can be scheduled reasonably and  the overall training time is 75 seconds, which is the largest training time of the three stages. However, when we use single node without 3-stage pipeline,  the three stages are executed sequentially and the training time of SFCTR is the summation of individual training time taken in the three stages and is around 150 seconds.

\section{Conclusion}
In this paper, we propose SFCTR, a MixCache-based distributed training system for CTR models with huge embedding table. SFCTR, which consists of Data-Loader, Host-Manager and GPU-Worker, overcomes the limitations of state-of-the-art distributed training systems for CTR prediction and achieves better efficiency. Firstly, SFCTR utilizes Virtual Sparse Id operation to reduce the amount of parameters/gradients transfer between Host-GPU and GPU-GPU. Secondly, SFCTR adopts MixCache mechanism to eliminate the latency of parameters/gradients transfer between Host-GPU. Thirdly, SFCTR uses a 3-stage pipeline that overlaps Data-Loader, Host-Manager and GPU-Worker. We conduct extensive experiments and ablation studies to demonstrate the superiority of SFCTR.

There are two interesting directions for future work. One is to improve the communication efficiency (such as All2All), which will further accelerate SFCTR. The other is to investigate the fast optimization method to speed up the convergence of deep CTR model with huge embedding table.

 \nocite{langley00}
\newpage
\bibliography{reference}
\bibliographystyle{ACM-Reference-Format}

\end{document}